\documentclass[conference,a4paper]{IEEEtran}

\usepackage{amssymb,amsthm}
\usepackage{amsfonts,amsmath}
\usepackage{latexsym}
\usepackage[mathscr]{euscript}
\usepackage{graphics}
\usepackage{graphicx}
\usepackage{cite}
\usepackage{cases}

\newcommand{\beq}{\begin{equation}}
\newcommand{\eeq}{\end{equation}}

\newcommand{\beqq}{\begin{equation*}}
\newcommand{\eeqq}{\end{equation*}}
\newcommand{\ei}{\end{itemize}}
\newcommand{\bi}{\begin{itemize}}

\newtheorem{prop}{Proposition}

\newtheorem{lemma}{Lemma}

\theoremstyle{remark}

\newcommand{\argmax}[1]{\arg{\hbox{$\underset{#1}{\max}\,$}}}

\begin{document}

\title{A Game Theoretic Analysis for Energy Efficient Heterogeneous Networks}

\author{Majed~Haddad\authorrefmark{1},~Piotr~Wiecek\authorrefmark{2},~Oussama~Habachi
\authorrefmark{3} and~Yezekael~Hayel\authorrefmark{3}\\
\authorrefmark{2}INRIA Sophia-Antipolis, Sophia-Antipolis, France\\
\authorrefmark{1}Institute of Mathematics and Computer Science, Wroclaw University of Technology, Poland\\
\authorrefmark{3}CERI/LIA, University of Avignon, Avignon, France
}

\maketitle
%\vspace{-1cm}
\begin{abstract}

Smooth and green future extension/scalability (e.g., from sparse to dense, from small-area dense to large-area dense, or from normal-dense to super-dense) is an important issue in heterogeneous networks. In this paper, we study energy efficiency of heterogeneous networks for both sparse and dense two-tier small cell deployments. We formulate the problem
as a hierarchical (Stackelberg) game in which the macro cell is the leader whereas the small cell is the follower. Both players want to strategically decide on their power allocation policies in order to maximize the energy efficiency of their
registered users. 
A backward induction method has been used to obtain a closed-form expression of the Stackelberg equilibrium. It is shown that the energy efficiency is maximized when only one sub-band is exploited for the players of the game depending on their fading channel gains. Simulation
results are presented to show the effectiveness of the proposed scheme.
\end{abstract}

\vspace{0.5cm}

\section{Introduction}

\begin{figure}[t]
\centering
\vspace{-7cm}
\hspace*{-4.2cm}
\includegraphics[height = 24cm,width=16.5cm]{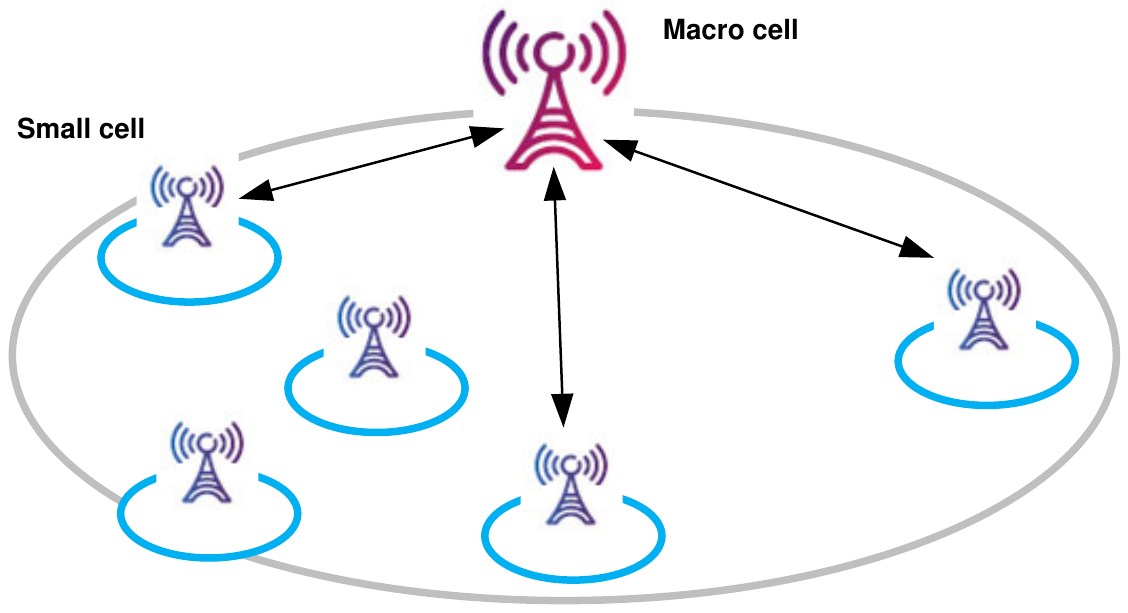}
\vspace{-11cm}
\caption{A heterogeneous networks with multi-layered systems of overlapping macro and small cells.}
\label{fig:sys}
\end{figure}

Small cells are deployed over the existing macro cell
network and share the same frequency spectrum with
macro cells (see Fig. \ref{fig:sys}). Due to spectral scarcity, the small cells and
macro cells have to reuse the allocated frequency band
partially or totally which leads to co-tier or cross-tier
interference \cite{andrews2012femtocells}.
Interference mitigation between neighboring small cells and between the macro cell and the small cell is considered to be one of the major challenges in
heterogeneous networks. Further, the conventional
radio resource management techniques for hierarchical cellular
system is not suitable for heterogeneous networks since the position
of the small cells is random depending on the users' service
requirement \cite{OFDMA:femtocell:Ekram12}, \cite{guruacharya:Stack:Hetnets2010}.
Several existing
works addressed the challenges of interference management in
heterogeneous networks \cite{Femto08Survey}, \cite{Hoydis2011SmallCells}. Moreover, for mobility connectivity performance, both sparse and dense deployments should be considered with equal priority as suggested in the 3GPP Release 12 \cite{3GPP-R12}.
Moreover, as the number of radio nodes
increases, backhaul becomes more important. Backhaul performance not only affects the data throughput available to users, but also the overall performance of the radio-access network \cite{EricssonWP12}. Recently, millimeter-wave \cite{mmWave_jsac09} was proposed as a potential candidate for the backhaul link as it enables multi-Gbps
wireless data communications. However, achieving these goals is usually at the expense of higher energy consumption. Therefore how to reduce power consumption while
satisfying the system throughput requirement becomes a vital
task in heterogeneous networks.

In this paper, we introduce a novel game-theoretic
framework in heterogeneous network which enables both the small cells and the macro cell to strategically decide on their
downlink power control policies. Due to the nature of heterogeneous networks architecture, we formulate the problem as a Stackelberg (hierarchical) game in which the
macro cell and the small cells strategically optimize the energy efficiency of their users.

\section{System Model}\label{sec:model}

Small cells
are controlled by a Home eNB (HeNB) whereas macro cells are controlled by a macro eNodeB (MeNB). Let $g_{f}^k$ denote the downlink channel gain between small cell user (SUE) $f$ and its serving HeNB on carrier $k$ with $k=1,\ldots,K$.
The macro cell users (MUEs) exist indoor as well and are served by the macro cell, where we denote by $g_{0}^k$ the downlink channel gain of a macro-user served by the macro cell on carrier $k$.
In our model, channel gain includes Rayleigh fading. Both the small cells and the macro cell
operate in a shared spectrum environment which gives rise to a mutual interference. $h_{0}^k$, resp. $h_{f}^k$, stands for the interfering signal from the macro cell eNodeB $0$, resp. the small cell Home eNodeB $f$, on carrier $k$. For realistic reasons we will assume that $K\geq N$ where $N$ stands for the number of players in the game, i.e., one macro cell (referred to hereafter and interchangeably as the leader) and $F$ small cells (referred to hereafter and interchangeably as followers).

\section{Network energy-efficiency analysis}
\label{sec:net-energy-eff}

Energy consumption can be decomposed into
the fixed-energy-consumption part and the dynamic-energy consumption
part \cite{EE_mag_11}. Fixed energy consumption, e.g., circuit
energy consumption, is the baseline energy consumed. Circuit energy consumption usually depends on both the hardware and software configurations of a device and is independent of the number of occupied channels.
In this paper, we will focus on the dynamic energy consumption which accounts for the transmission
energy consumed in radio frequency transmission circuits depending
on the number of occupied channels. The system model adopted throughout the paper is based
on the seminal paper \cite{goodman-pcomm-2000} that defines the energy efficiency
framework.
We consider a utility function that allows one to measure
the corresponding trade-off between the transmission benefit
(total throughput over the $K$ carriers) and the cost (total power over the $K$ carriers):
\beq\label{eq:util-mc}
u_n(\mathbf{p_1},\ldots,\mathbf{p_{N}})=\frac{\displaystyle R_n \cdot \sum_{k=1}^K  f(\gamma_{n}^k)}{\displaystyle\sum_{k=1}^K p_{n}^k}
\eeq
where $\mathbf{p_{n}}$ is the power allocation
vector of user  $n$ over all carriers $k$, i.e., $\mathbf{p_{n}}=(p_n^1,\ldots,p_n^K)$.
$R_n$ and $\gamma_n^k$ are respectively the transmission
rate and the SINR of user $n$ over carrier $k$. $f(\cdot)$ is the "efficiency" function which measures the
packet success rate. The utility function $u_n$, that has bits per joule as units, perfectly captures the trade-off between throughput, and battery life and is
particularly suitable for applications where energy efficiency is
crucial.
The efficiency function $f(\cdot)$ is an increasing, continuous sigmoidal function. It can be
shown that for a sigmoidal efficiency function, the utility function in (\ref{eq:util-mc}) is a
quasi-concave function of the user's transmit power \cite{rodriguez-globecom-2003}, and we will use this assumption throughout our paper.
Clearly, a message from Eq.~\ref{eq:util-mc} in terms of power usage is that a device should not exploit, in general, all the available radiated power at the transmitter to maximize its energy efficiency.

\section{The game theoretic formulation}\label{sec:game-model}
\subsection{The non-cooperative game problem}\label{sec:nash-model}

An important solution concept of the game under consideration is the Nash equilibrium (NE)
\cite{nash50}, which is a fundamental concept in non-cooperative strategic games. It is a vector
of strategies $\mathbf{p}^{NE} = \{\mathbf{p_1}^{NE},\ldots,\mathbf{p_N}^{NE}\}$, one for
each player, such that no player has incentive to unilaterally deviate, i.e.,
$u_n(\mathbf{p_n}^{NE},\mathbf{p_{-n}}^{NE})\geq u_n(\mathbf{p_n},\mathbf{p_{-n}}^{NE})$ for every action $\mathbf{p_n} \neq
\mathbf{p_n}^{NE}$, where the $-n$ subscript on vector $\mathbf{p}$ stands for "except user $n$", i.e.,
$\mathbf{p_{-n}} = \left\{\mathbf{p_1},\ldots, \mathbf{p_{n-1}}, \mathbf{p_{n+1}}, \ldots , \mathbf{p_{N}} \right\}$.

\subsection{The hierarchical game formulation}
\label{sec:stack-model}

There are many motivations for studying wireless
networks with hierarchical structures, but the most important
ones are to improve the network efficiency and modeling
aspect. The Stackelberg game has been first proposed in economics and independently in
biology for modeling optimal behavior against nature \cite{Tirole91}. It is also a natural setting for heterogeneous wireless networks due to the absence
of coordination among the small cells and between small cells and macro cells. At the core lies the idea that the utility
of the leader obtained at the Stackelberg equilibrium can often be improved over his utility obtained at
the Nash equilibrium when the two users play simultaneously. It has been proved in \cite{samson-twc09} that when only one carrier is available for the players this result is true for both the leader and the follower. The goal is then to find a Stackelberg
equilibrium in this two-step game.
It is noteworthy that if there exists a Nash equilibrium in a game, there exists at most one
Stackelberg equilibrium \cite{Tirole91}.
In this work, we consider a Stackelberg game framework in which the macro cell (or the leader) decides first his power allocation
vector $\mathbf{p_{0}}=(p_0^1,\ldots,p_0^K)$ and based on this value, a small cell (or follower) $f$ will adapt its power
allocation vectors  $\mathbf{p_{f}}=(p_f^1,\ldots,p_f^K)$ for $f=1,\dots,F$.
A Stackelberg equilibrium can then be determined using a bi-level approach, where, given the action
of the leader, we compute the best-response function of the follower (the function
$\overline{p}_f(\cdot)$ for $f=1,\dots,F$) and find the actions of the followers which maximize their utilities.

\section{Sparse Network Model}\label{sec:sparse}

Two-tier heterogeneous networks have
been proposed to accommodate the rapid increase in wireless data traffic
in indoor environments. In this section, we first consider a wireless network where small cells are
deployed in a sparse manner to cover the hotspots and are
overlaid by a single macro cell. In such environment, macro eNB
will have very good channel conditions to its macro cell users while
signals received from the outdoor small cell users will be highly
attenuated by the macro cell transmission. We will thus ignore interference from small cells transmissions since their
contribution is minimal in a sparse network yielding the following MUE and SUE SINR on carrier $k$ respectively

\begin{eqnarray}
\displaystyle \gamma_{0}^k &=&\frac{g_{0}^k p_{0}^k}{\sigma^2}\\
\label{eq:gamma-fol-sparse}\displaystyle \gamma_{f}^k&=&\frac{g_{f}^k p_{f}^k}{\sigma^2+ {h_{0}^k} p_{0}^k }, \qquad \text{for}\,\,\, f=1,\dots,F
\end{eqnarray}
where $\sigma^2$ stands for the noise variance.
Here and in the sequel $B_n$, $n=0,1,\ldots,F$ denotes the carrier for which player $n$'s signal channel gain $g_n^k$ is the biggest, while $S_n$, $n=0,1,\ldots,F$ denotes the carrier for which his signal channel gain $g_n^k$ is the second biggest. The following proposition will be true in this situation.
\begin{prop}
\label{prop:downlink}
Let $\gamma^*$ be the unique positive solution to the equation
\begin{equation}\label{eq:gamma*}
x\,f^{\prime}(x)=f(x)
\end{equation}
In the sparse network model, the equilibrium power allocations of each of the players are as follows:\\
The power allocation of the leader is:
$$\overline{p_0^k}=\left\{ \begin{array}{ll}
  \frac{\gamma^*\sigma^2}{g_0^k}&\mbox{for }k=B_0\\
  0&\mbox{otherwise}
  \end{array}\right.$$
The power allocation of the follower $f$ if $B_f\neq B_0$ is
$$\overline{p_f^k}=\left\{ \begin{array}{ll}
  \frac{\gamma^*\sigma^2}{g_f^k}&\mbox{for}k=B_f\\
  0&\mbox{otherwise}
  \end{array}\right.$$
The power allocation of the follower $f$ if $B_f=B_0$ and $\frac{g_f^{B_f}}{g_f^{S_f}}\geq 1+\frac{h_0^{B_f}}{g_0^{B_f}}\gamma^*$ is
$$\overline{p_f^k}=\left\{ \begin{array}{ll}
  \frac{\gamma^*\sigma^2(g_0^{k}+\gamma^*h_0^{k})}{g_0^kg_f^k}&\mbox{for }k=B_f\\
  0&\mbox{otherwise}
  \end{array}\right.$$
The power allocation of the follower $f$ if $B_f=B_0$ and $\frac{g_f^{B_f}}{g_f^{S_f}}\leq 1+\frac{h_0^{B_f}}{g_0^{B_f}}\gamma^*$ is
$$\overline{p_f^k}=\left\{ \begin{array}{ll}
  \frac{\gamma^*\sigma^2}{g_f^k}&\mbox{for }k=S_f\\
  0&\mbox{otherwise}
  \end{array}\right.$$
\end{prop}

Prop.~\ref{prop:downlink} proved that the best-response of two actors of the system is to use only one carrier depending on their fading channel gains.
This result differs from what it is usually obtained by throughput-based systems. Indeed, it is well known that maximizing the sum throughput leads to a water-filling solution \cite{Goldsmith97capacityof} where only a certain number of bands are exploited depending on the channel gains. In particular, when the SNR is low (resp. high), only one (resp. every) band is exploited. Energy efficiency, on the other hand, is always maximized by using a single band. This means that each of MeNB and HeNB always leaves $K-1$ bands completely free.

\section{Dense Network Model}\label{sec:dense}

Consider now a scenario in which a lot of small cells are densely deployed to support huge traffic over a relatively wide area covered by the macro cell. This is typically suitable for dense urban networks or a large shopping mall where the coverage of the small cell layer is generally discontinuous between different hotspot areas. System simulations in \cite{HeNB-3GPP} have suggested that this scenario should require special attention since it creates a significant interference impact. The impact of this interference increases as the density of small cells within the macro cell coverage
area increases. We will focus on mitigating the cross-tier interference between the macro cell and small cells in a two-tier downlink model.

More specifically, we consider a dense two-tier network with $f$ small cells overlaid by a single macro cell giving rise to non-negligible interference from small cells. The resulting SINR of MUE $0$ served by the macro cell on carrier $k$ is given by

\beq\label{eq:2tier}
\displaystyle \gamma_{0}^k=\frac{g_{0}^k p_{0}^k}{\sigma^2+ \displaystyle\sum_{\substack{f=1}}^F {h_{f}^k} p_{f}^k}
\eeq

On the other hand, downlink transmissions of a small cell Home eNB suffer from interference with the macro cell transmissions yielding the same SINR expression in (\ref{eq:gamma-fol-sparse}).

In this section we present an algorithm for finding exact equilibrium in the partial interference two-tier system model. In the whole section we make the following two additional assumptions about function $f$:\\
{\bf (A1)} $f^{\prime}(0^+)=0$ or $f^{\prime}(0^+)>0$ and $\frac{f^{{\prime}{\prime}}(0^+)}{f^{\prime}(0^+)}>2\gamma^*\max_{k\leq K}\left[\frac{h_0^k}{g_0^k}\sum_{f=1}^F\frac{h_f^k}{g_f^k}\right]$.\\
{\bf (A2)} For any $a>0$ the equation
$$(x-ax^2)f^{\prime}(x)=f(x)$$
always has exactly one positive solution.\\
These two assumptions assure that the game under consideration always has an equilibrium and that it can be found using the algorithm presented in the proposition below. Note however that in particular, for the most standard form of $f$,
$$f(x)=(1- e^{-x})^M,$$ both assumptions are satisfied, so they are not overly problematic.
\begin{prop}
\label{prop:uplink_sparse}
Let $\gamma^*$ be defined as in (\ref{eq:gamma*})
The following procedure gives the equilibrium power allocations for all of the players in the two-tier model:
\begin{enumerate}
\item Let every follower $f$ choose his best carrier $B_f$. For each $k$ denote by $F(k)$ the number of followers who choose $k$ and sort these followers in decreasing order by their $\theta_f=\frac{g_f^{B_f}}{g_f^{S_f}}$. Denote by $f(k,l)$ the follower choosing carrier $k$ with $l$-th biggest $\theta_f$.
\item For every $k$ and each $L\leq F(k)$ let $\eta_{k,L}=\sum_{l=1}^L\frac{h_{f(k,l)}^k}{g_{f(k,l)}^k}$ and find the positive solution to the equation
$$(x-\frac{h_0^k\gamma^*\eta_{k,L}}{g_0^k}x^2)f'(x)=f(x).$$
Let $\gamma^{**}_{k,L}$ be this solution. %the solution giving the biggest value of $\frac{f(x)(g_0^k-h_0^k\gamma^*\eta_{k,L}x)}{x}$.
\item For each $k$ let $F^*(k)$ be the biggest $L$ such that $$g_{f(k,l)}^k(g_0^k-\gamma^{**}_{k,l}\gamma^*\eta_{k,l}h_0^k)>g_{f(k,l)}^{S_{f(k,l)}}(g_0^k+h_0^k\gamma^{**}_{k,l})$$
and for each $l\leq F^*(k)$ compute the four values:
$$V_0^k(l)=\frac{f(\gamma^{**}_{k,l})(g_0^k-\gamma^{**}_{k,l}\gamma^*\eta_{k,l}h_0^k)R_0}{\gamma^{**}_{k,l}(1+\gamma^*\eta_{k,l})\sigma^2},$$
$$\widetilde{p}_0^k(l)=\frac{\gamma^{**}_{k,l}(1+\gamma^*\eta_{k,l})\sigma^2}{g_0^k-\gamma^*\gamma^{**}_{k,l}\eta_{k,l}h_0^k},$$
$$\widehat{V}_0^k(l)=\frac{f(\frac{g_0^k\widehat{p}_0^k(l)}{\sigma^2(1+\gamma^*\eta_{k,l-1})+\gamma^*\eta_{k,l-1}h_0^k\widehat{p}_0^k(l)})}{\widehat{p}_0^k(l)},$$
$$\widehat{p}_0^k(l)=\frac{\sigma^2(g_{f(k,l)}^k-g_{f(k,l)}^{S_{f(k,l)}})}{h_0^kg_{f(k,l)}^{S_{f(k,l)}}}.$$
\item For $l< F^*(k)$ do the following steps:\\
If $\widetilde{p}_0^k(l)<\widehat{p}_0^k(l+1)$, put $V_0^k(l)=\widehat{V}_0^k(l+1)$ and $\widetilde{p}_0^k(l)=\widehat{p}_0^k(l+1)$.\\
If $\widetilde{p}_0^k(l)>\widehat{p}_0^k(l)$, put $V_0^k(l)=\widehat{V}_0^k(l)$ and $\widetilde{p}_0^k(l)=\widehat{p}_0^k(l)$.
\item Find $(\widehat{k},\widehat{l})$, maximizing $V_0^k(l)$ for $l\leq F^*(k)$, $k=1,\ldots,K$. Take $F^*(\widehat{k})=\widehat{l}$,
$\widetilde{p}_0^{\widehat{k}}=\widetilde{p}_0^{\widehat{k}}(\widehat{l})$ and $\widetilde{p}_{f(\widehat{k},l)}^{\widehat{k}}=\frac{\gamma^*(\sigma^2+h_0^{\widehat{k}}\widetilde{p}_0^{\widehat{k}}(\widehat{l}))}{g_{f(\widehat{k},l)}^{\widehat{k}}}$ for $l=1,\ldots,\widehat{l}$.
\end{enumerate}
The equilibrium power allocations of the leader are defined by:
$$\overline{p}_0^k=\left\{ \begin{array}{ll}
  \widetilde{p}_0^k&\mbox{for }k=\widehat{k}\\
  0&\mbox{otherwise}
  \end{array}\right.$$
The equilibrium power allocations of follower $f$ when $f=f(\widehat{k},l)$ with $l\leq F^*(\widehat{k})$ are defined by:
$$\overline{p}_{f(k,l)}^k=\left\{ \begin{array}{ll}
  \widetilde{p}_{f(k,l)}^k&\mbox{for }k=\widehat{k}\\
  0&\mbox{otherwise}
  \end{array}\right.$$
The equilibrium power allocations of follower $f$ when $f=f(\widehat{k},l)$ with $l> F^*(\widehat{k})$ are defined by:
$$\overline{p}_{f(k,l)}^k=\left\{ \begin{array}{ll}
  \frac{\gamma^*\sigma^2}{g_f^k}&\mbox{for }k=S_f\\
  0&\mbox{otherwise}
  \end{array}\right.$$
Finally, the equilibrium power allocations of follower $f$ when $B_f\neq \widehat{k}$ are defined by:
$$\overline{p}_{f(k,l)}^k=\left\{ \begin{array}{ll}
  \frac{\gamma^*\sigma^2}{g_f^k}&\mbox{for }k=B_f\\
  0&\mbox{otherwise}
  \end{array}\right.$$
\end{prop}

Although the formulation of Prop.~\ref{prop:uplink_sparse} is rather complicated, we end up having similar observations than for Prop.~\ref{prop:downlink}.\\

\section{Numerical Illustrations}\label{sec:simul}

We present a scenario with small cells overlaying an existing macro cell network. We first provide a performance comparison of the proposed Stackelberg model described in Section \ref{sec:dense} and the following \emph{traditional} communication schemes:
\begin{itemize}
\item \emph{the Nash model}: both the macro cell and the small cell choose their power level according to \cite{meshkati-jsac-2006} in a non-cooperative manner,
\item \emph{the best channel model}: all players choose to transmit on their "best" channels.\\
\end{itemize}

\begin{figure}[t]
\centering
\vspace{-6cm}
\hspace*{-2cm}
\includegraphics[height = 17cm,width=13cm]{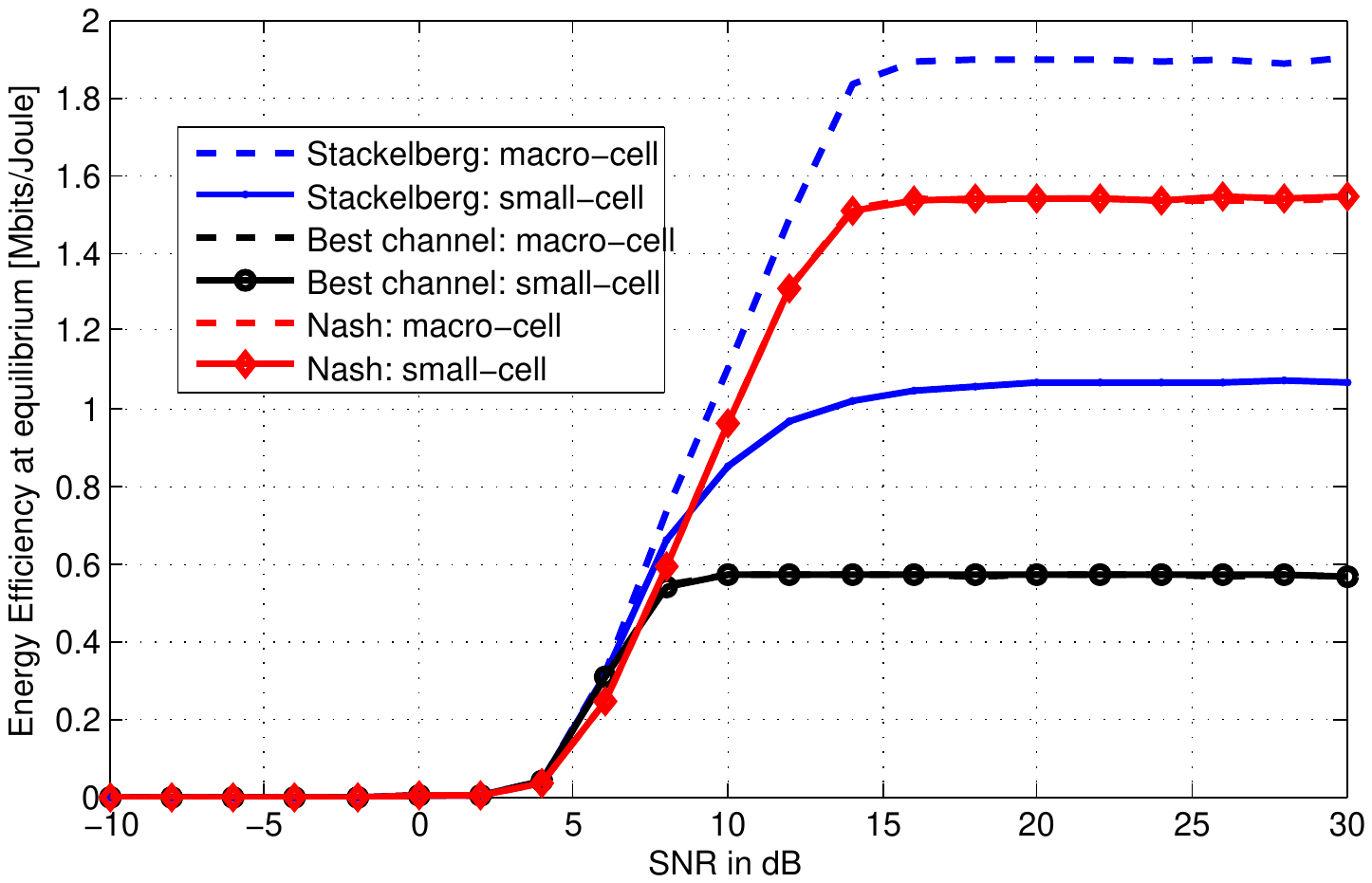}
\vspace{-6cm}
\caption{Energy efficiency at the equilibrium as function of the SNR for different schemes with $5$ carriers.}
\label{fig:eff}
\end{figure}

As expected, results in Figure \ref{fig:eff} show that the MeNB (i.e., the leader) performs better at Stackelberg than for the other strategies in terms of energy efficiency. This is at the expense of the HeNB (i.e., the follower) which performs worse at Nash equilibrium. This is due to the fact that in Nash model, the MeNB does not anticipate the HeNB's action like in the Stackelberg model.\\

\begin{figure}[t]
\vspace{-4cm}
\hspace*{-1cm}
\includegraphics[height = 14cm,width=11cm]{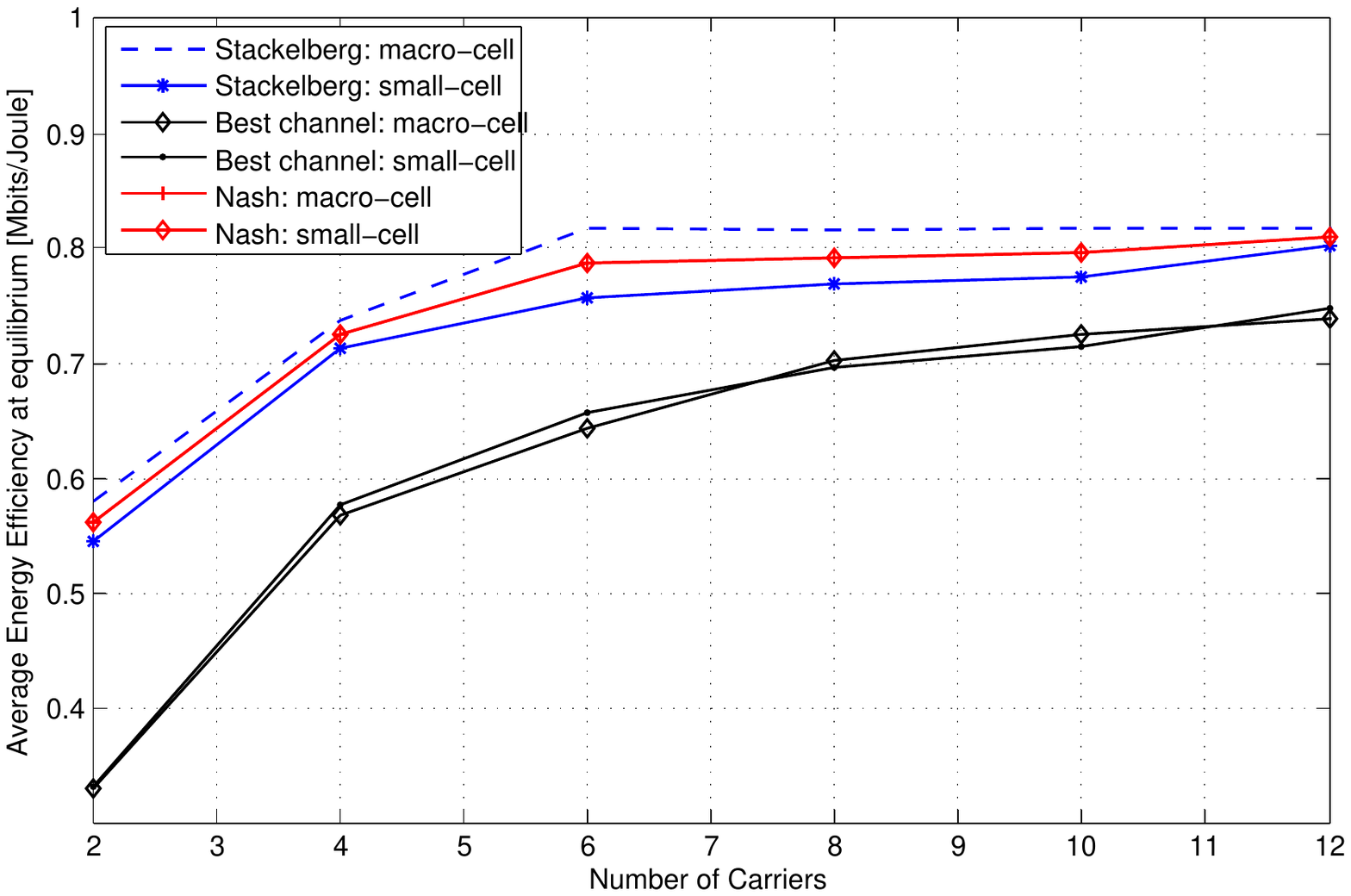}
\vspace{-4.5cm}
\caption{Energy efficiency at the equilibrium as function of the number of carriers.}
\label{fig:utility}
\end{figure}

In Figure \ref{fig:utility}, we plot the energy efficiency at equilibrium as a function of the number of carriers. Again, we observe that the MeNB of the Stackelberg scheme achieves the best energy efficiency compared to the other schemes. Moreover, we found out that both the utility of the leader and the follower are increasing with the number of carriers. As the latter increases, all configurations tend towards having the same energy efficiency since the channel diversity gain tends to $0$.

\section{Conclusion}\label{sec:conc}

In this paper, we have introduced a novel game-theoretic
framework in heterogeneous network which enables both the small cells and the macro cell to strategically decide on their downlink power control policies. Due to the nature of heterogeneous networks architecture, we have formulated a hierarchical game in which the macro cell and the small cells strategically optimize the energy efficiency of their users. We have derived analytically the equilibrium for the sparse and two-tier dense network. In particular, we have shown that the energy efficiency is maximized when only one sub-band is exploited for the players of the game and the other sub-bands are left unused. This result differs from what it is usually obtained by throughput-based systems where only a certain number of bands are exploited depending on the channel gains.
Simulation results assessed the performance of the proposed
approach in various settings.

%\newpage
\bibliographystyle{IEEEtran}
%\bibliography{mybib}
\bibliography{C:/Users/mhaddad/Dropbox/mybib}

\appendix
Before we present the proofs of our main results, we cite an important result from \cite{meshkati-jsac-2006} as a lemma:

\begin{lemma}
\label{prop:follower-power}
Given the power allocation vector $\mathbf{p_0}$ of the leader, the best-response of the follower $f$ for $f=1,\dots,F$ is
given by
\begin{equation}\label{eq:follower}
\overline{p}_f^k(\mathbf{p_0})= \left\{\begin{array}{lr}\displaystyle
\frac{\gamma^{*}(\sigma^2+g_{0}^k p_{0}^k)}{g_{f}^k},&\quad \mbox{for} \quad k = L_f(\mathbf{p_0}),\\
0,& \quad \mbox{for all}\quad  k \neq L_f(\mathbf{p_0})
\end{array}
\right.
\end{equation}
with $L_f(\mathbf{p_0})=\argmax{k} \frac{g_f^k}{\sigma^2+h_0^kp_0^k}$ and $\gamma^*$ is the unique
(positive) solution of the equation
(\ref{eq:gamma*}).
\end{lemma}

\subsection{Proof of Proposition \ref{prop:downlink}}
\begin{IEEEproof}
Since the macro cell user does not experience any interference, his utility from using carrier $k$ is
\beq
\label{leader_utility_uds}
\frac{f(\frac{g_0^{k}p_0^{k}}{\sigma^2})}{p_0^{k}},
\eeq
which is by Lemma \ref{prop:follower-power} maximized for $p_0^k=\frac{\gamma^*\sigma^2}{g_0^k}$. If we substitute it into (\ref{leader_utility_uds}), we obtain that his utility from carrier $k$ is $\frac{f(\gamma^*)g_0^k}{\gamma^*\sigma^2}$, which is obviously maximized for $k=B_0$.

Now, knowing that the leader will use carrier $B_0$ and power $p_0^{B_0}=\frac{\gamma^*\sigma^2}{g_0^{B_0}}$, follower $f$, experiencing interference only with the leader may either choose carrier $B_0$, obtaining as utility
$$\frac{f(\frac{g_f^{B_0}p_f^{B_0}}{\sigma^2+h_0^{B_0}p_0^{B_0}})}{p_f^{B_0}},$$
or choose some other carrier $k$ where his utility will be
$$\frac{f(\frac{g_f^{k}p_f^{k}}{\sigma^2})}{p_f^{k}}.$$
The former is (again by Lemma \ref{prop:follower-power}) maximized for $p_f^{B_0}=\frac{\gamma^*(\sigma^2+h_0^{B_0}p_0^{B_0})}{g_f^{B_0}}=\frac{\gamma^*\sigma^2(g_0^{B_0}+\gamma^*h_0^{B_0})}{g_0^{B_0}g_f^{B_0}}$ and then equal to
\beq
\label{follower_utility_uds}
\frac{f(\gamma^*)g_f^{B_0}}{\gamma^*(\sigma^2+h_0^{B_0}p_0^{B_0})}=\frac{f(\gamma^*)g_f^{B_0}}{\gamma^*\sigma^2(1+\frac{h_0^{B_0}}{g_0^{B_0}}\gamma^*)},
\eeq
while the latter for $p_f^k=\frac{\gamma^*\sigma^2}{g_f^k}$ and then equal to $\frac{f(\gamma^*)g_f^k}{\gamma^*\sigma^2}$. Obviously, if $g_f^k\geq g_f^{B_0}$ the latter is bigger, and so for $B_f\neq B_0$ $f$ chooses to allocate to carrier $B_f$ power  $p_f^{B_f}=\frac{\gamma^*\sigma^2}{g_f^{B_f}}$. Otherwise (when $B_f=B_0$) he compares the value (\ref{follower_utility_uds}) he may obtain on $B_f$ and the one he can obtain on the best of his remaining carriers, $S_f$,
$\frac{f(\gamma^*)g_f^{S_f}}{\gamma^*\sigma^2}$. Choosing the bigger one gives the result given in the proposition.
\end{IEEEproof}

%\vspace{-0.5cm}
\subsection{Proof of Proposition \ref{prop:uplink_sparse}}
\begin{IEEEproof}
We first show that the players will not have incentive to change their power allocations defined by the algorithm, without changing the carriers they use. Note that, since small cells experience interference only with the macro cell $0$, they always only adjust their power to that chosen by the macro cell $0$, according to Lemma \ref{prop:follower-power}. This is done in every case considered in the algorithm.

Now suppose that $\widehat{k}$ and $F^*(\widehat{k})>0$ are chosen by the algorithm and that step 4) did not change the value of $V_0^k$. We will show that in this case player $0$ will not have incentive to change his power. There are $F^*(\widehat{k})$ followers using carrier $\widehat{k}$, and these are players $f(\widehat{k},l)$, $l\leq F^*(\widehat{k})$.
The leader's SINR is thus of the form:
$$
\gamma_0^{\widehat{k}}=\frac{g_0^{\widehat{k}}p_0^{\widehat{k}}}{\sigma^2+\sum_{l=1}^{F^*(\widehat{k})}h_{f({\widehat{k}},l)}^{\widehat{k}}p_{f({\widehat{k}},l)}^{\widehat{k}}}
$$
If we substitute the power used by the followers by Lemma \ref{prop:follower-power} into it, we obtain
\begin{eqnarray}
\gamma_0^{\widehat{k}}&=&
\frac{g_0^{\widehat{k}}p_0^{\widehat{k}}}{\sigma^2+\sum_{l=1}^{F^*(\widehat{k})}h_{f({\widehat{k}},l)}^{\widehat{k}}\frac{\gamma^*(\sigma^2+h_0^{\widehat{k}}p_0^{\widehat{k}})}{g^{\widehat{k}}_{f({\widehat{k}},l)}}}\nonumber\\
&=&\frac{g_0^{\widehat{k}}p_0^{\widehat{k}}}{\sigma^2(1+\eta_{{\widehat{k}},F^*({\widehat{k}})}\gamma^*)+\eta_{{\widehat{k}},F^*({\widehat{k}})}\gamma^*h^{\widehat{k}}_0p^{\widehat{k}}_0}\nonumber\\
&=&\frac{g_0^{\widehat{k}}}{\eta_{{\widehat{k}},F^*({\widehat{k}})}\gamma^*h_0^{\widehat{k}}}\left(1-\frac{1}{1+\frac{p_0^{\widehat{k}}\eta_{{\widehat{k}},F^*({\widehat{k}})}\gamma^*h_0^{\widehat{k}}}{\sigma^2(1+\eta_{{\widehat{k}},F^*({\widehat{k}})}\gamma^*)}}\right)
\label{sinr_transformed0}
\end{eqnarray}
where $\eta_{{\widehat{k}},F^*({\widehat{k}})}=\sum_{l=1}^{F^*({\widehat{k}})}\frac{h^{\widehat{k}}_{f({\widehat{k}},l)}}{g^{\widehat{k}}_{f({\widehat{k}},l)}}$.
Now we differentiate $\gamma_0^{\widehat{k}}$ with respect to $p_0^{\widehat{k}}$:
\begin{eqnarray}
\frac{\partial \gamma_0^{\widehat{k}}}{\partial p_0^{\widehat{k}}}&=&\frac{g_0^{\widehat{k}}}{\sigma^2(1+\eta_{\widehat{k},F^*(\widehat{k})}\gamma^*)(1+\frac{p_0^{\widehat{k}}\eta_{{\widehat{k}},F^*({\widehat{k}})}\gamma^*h_0^{\widehat{k}}}{\sigma^2(1+\eta_{{\widehat{k}},F^*({\widehat{k}})}\gamma^*)})^2}\\\nonumber
&=&\frac{g_0^{\widehat{k}}\sigma^2(1+\eta_{\widehat{k},F^*(\widehat{k})}\gamma^*)}{{(\sigma^2(1+\eta_{\widehat{k},F^*(\widehat{k})}\gamma^*)+\eta_{\widehat{k},F^*(\widehat{k})}\gamma^*h_0^{\widehat{k}}p_0^{\widehat{k}})^2}} \\\nonumber
&=&\frac{1}{p_0^{\widehat{k}}}\frac{\sigma^2(1+\eta_{\widehat{k},F^*(\widehat{k})}\gamma^*)}{g_0^{\widehat{k}}p_0^{\widehat{k}}}(\gamma_0^{\widehat{k}})^2
\label{sinr_diff1}
\end{eqnarray}

Next we can transform (\ref{sinr_transformed0}) into
$$p_0^{\widehat{k}}=\frac{\sigma^2(1+\eta_{\widehat{k},F^*(\widehat{k})}\gamma^*)\gamma_0^{\widehat{k}}}{g_0^{\widehat{k}}-\eta_{\widehat{k},F^*(\widehat{k})}\gamma^*h_0^{\widehat{k}}\gamma_0^{\widehat{k}}}$$
and substitute it into (\ref{sinr_diff1}), obtaining
\beq
\label{sinr_diff_trans}
\frac{\partial \gamma_0^{\widehat{k}}}{\partial p_0^{\widehat{k}}}=\gamma_0^{\widehat{k}}(1-\eta_{\widehat{k},F^*(\widehat{k})}\gamma^*\frac{h_0^{\widehat{k}}}{g_0^{\widehat{k}}}\gamma_0^{\widehat{k}}).
\eeq

The utility of the leader is
\beq
\label{leader_utility0}
\frac{\gamma_0^{\widehat{k}}}{p_0^{\widehat{k}}}=\frac{f(\frac{g_0^{\widehat{k}}p_0^{\widehat{k}}}{\sigma^2(1+\eta_{{\widehat{k}},F^*({\widehat{k}})}\gamma^*)+\eta_{{\widehat{k}},F^*({\widehat{k}})}\gamma^*h^{\widehat{k}}_0p^{\widehat{k}}_0})}{p_0^{\widehat{k}}}.
\eeq
The first order condition for the maximization of (\ref{leader_utility0}) is
$$0=\frac{\partial(\frac{R_0f(\gamma_0^{\widehat{k}})}{p_0^{\widehat{k}}})}{\partial p_0^{\widehat{k}}}=R_0\frac{-f(\gamma_0^{\widehat{k}})+f^{\prime}(\gamma_0^{\widehat{k}})\frac{\partial \gamma_0^{\widehat{k}}}{\partial p_0^{\widehat{k}}}p_0^{\widehat{k}}}{(p_0^{\widehat{k}})^2}.$$
If we substitute (\ref{sinr_diff_trans}) into it we obtain
\beq
\label{this_eq2}
(\gamma_0^{\widehat{k}}-\frac{h_0^k\gamma^*\eta_{{\widehat{k}},F^*({\widehat{k}})}}{g_0^k}(\gamma_0^{\widehat{k}})^2)f'(\gamma_0^{\widehat{k}})=f(\gamma_0^{\widehat{k}}).
\eeq
By (A2) there is a unique solution to this equation. Moreover, this solution gives the maximum value of the macro cell's utility function, as (A1) implies that for $\gamma\rightarrow 0^+$ the utility function is increasing. Thus the value of
$$\widetilde{p}_0^{\widehat{k}}=\frac{\gamma^{**}_{{\widehat{k}},F^*({\widehat{k}})}(1+\gamma^*\eta_{{\widehat{k}},F^*({\widehat{k}})})\sigma^2}{g_0^{\widehat{k}}-\gamma^*\gamma^{**}_{{\widehat{k}},F^*({\widehat{k}})}\eta_{{\widehat{k}},F^*({\widehat{k}})}h_0^{\widehat{k}}}$$
associated with the solution to (\ref{this_eq2}), $\gamma^{**}_{{\widehat{k}},F^*({\widehat{k}})}$,
gives the biggest possible value of the macro cell's utility when $F^*(\widehat{k})$ small cells also use carrier $\widehat{k}$.
If however $\widetilde{p}_0^{\widehat{k}}$ defined above will be smaller than
$$\widehat{p}_0^{\widehat{k}}(F^*({\widehat{k}})+1)=\frac{\sigma^2(g_{f({\widehat{k}},F^*({\widehat{k}})+1)}^{\widehat{k}}-g_{f({\widehat{k}},F^*({\widehat{k}})+1)}^{S_{f({\widehat{k}},F^*({\widehat{k}})+1)}})}{h_0^{\widehat{k}}g_{f({\widehat{k}},F^*({\widehat{k}})+1)}^{S_{f({\widehat{k}},F^*({\widehat{k}})+1)}}},$$
small cell $f(\widehat{k},F^*({\widehat{k}})+1)$ will want to change his carrier to $\widehat{k}$. Thus in that case the value of $\widetilde{p}_0^{\widehat{k}}$ will have to be increased to the value of $\widehat{p}_0^{\widehat{k}}(F^*({\widehat{k}})+1)$ to avoid this situation. Similarly, when $\widetilde{p}_0^{\widehat{k}}$ will be bigger than
$$\widehat{p}_0^{\widehat{k}}(F^*({\widehat{k}}))=\frac{\sigma^2(g_{f({\widehat{k}},F^*({\widehat{k}}))}^{\widehat{k}}-g_{f({\widehat{k}},F^*({\widehat{k}}))}^{S_{f({\widehat{k}},F^*({\widehat{k}}))}})}{h_0^{\widehat{k}}g_{f({\widehat{k}},F^*({\widehat{k}}))}^{S_{f({\widehat{k}},F^*({\widehat{k}}))}}},$$
small cell $f(\widehat{k},F^*({\widehat{k}}))$ will want to change his carrier from $\widehat{k}$ to $S_{f(\widehat{k},F^*({\widehat{k}}))}$. Thus in that case the value of $\widetilde{p}_0^{\widehat{k}}$ will have to be decreased to the value of $\widehat{p}_0^{\widehat{k}}(F^*({\widehat{k}}))$ to avoid it.

Next we show that none of the players will have an incentive to change their carriers. For macro cell player $0$ this is obvious, because the algorithm chooses the carrier $\widehat{k}$ and the number of small cells transmitting on it $\widehat{l}$ in order to maximize his utility, so changing the carrier will obviously decrease his utility. As far as small cells whose best carrier is not $\widehat{k}$ are concerned -- each of them transmits on his best carrier obtaining his maximal possible utility, so none of them will have an incentive to change his carrier. Finally, let $f$ be a small cell whose $B_f=\widehat{k}$ such that $f=f(\widehat{k},l^*)$. If he transmits on carrier $\widehat{k}$, the condition
\begin{eqnarray}
g_{f(\widehat{k},F^*(\widehat{k}))}^{\widehat{k}}(g_0^{\widehat{k}}-\gamma^{**}_{{\widehat{k}},F^*(\widehat{k})}\gamma^*\eta_{{\widehat{k}},F^*(\widehat{k})}h_0^{\widehat{k}})
>\\ \nonumber
\qquad\qquad\qquad g_{f({\widehat{k}},F^*(\widehat{k}))}^{S_{f({\widehat{k}},F^*(\widehat{k}))}}(g_0^{\widehat{k}}+h_0^{\widehat{k}}\gamma^{**}_{{\widehat{k}},F^*(\widehat{k})}),
\end{eqnarray}
which is assured by the algorithm, implies that also
\begin{eqnarray}
\label{ineq_l}
g_{f(\widehat{k},l^*)}^{\widehat{k}}(g_0^{\widehat{k}}-\gamma^{**}_{{\widehat{k}},l^*}\gamma^*\eta_{{\widehat{k}},l^*}h_0^{\widehat{k}})>
%\\ \nonumber
%\qquad\qquad\qquad
g_{f({\widehat{k}},l^*)}^{S_{f({\widehat{k}},l^*)}}(g_0^{\widehat{k}}+h_0^{\widehat{k}}\gamma^{**}_{{\widehat{k}},l^*}),
\end{eqnarray}
as $l^*\leq F^*(\widehat{k})$ (because this is true for all the small cells transmitting on $\widehat{k}$) and thus
$$\frac{g_{f(\widehat{k},F^*(\widehat{k}))}^{B_{f(\widehat{k},F^*(\widehat{k}))}}}{g_{f(\widehat{k},F^*(\widehat{k}))}^{S_{f(\widehat{k},F^*(\widehat{k}))}}}=\frac{\widehat{k}}{g_{f(\widehat{k},F^*(\widehat{k}))}^{S_{f(\widehat{k},F^*(\widehat{k}))}}}\leq\frac{\widehat{k}}{S_{f(\widehat{k},l^*)}}=\frac{B_{f(\widehat{k},l^*)}}{S_{f(\widehat{k},l^*)}}$$
(because they are ordered by the algorithm in decreasing order by their $\frac{g^{B_f}_f}{g^{S_f}_f}$).
But (\ref{ineq_l}) can be rewritten as
\begin{eqnarray}\frac{f(\gamma^*)g^{\widehat{k}}_{f(\widehat{k},l^*)}(g_0^{\widehat{k}}-\gamma^{**}_{\widehat{k},F^*(\widehat{k})}\gamma^*\eta_{\widehat{k},F^*(\widehat{k})}h_0^{\widehat{k}})R_{f(\widehat{k},l^*)}}{\gamma^*\sigma^2(g_0^{\widehat{k}}+h_0^{\widehat{k}}\gamma^{**}_{\widehat{k},F^*(\widehat{k})})}>\\\nonumber
\frac{f(\gamma^*)g^{S_{f(\widehat{k},l^*)}}_{f(\widehat{k},l^*)}R_{f(\widehat{k},l^*)}}{\gamma^*\sigma^2}
\end{eqnarray}
But the LHS of the above inequality is $f$'s utility when he uses carrier $\widehat{k}$, while the RHS is his utility when he uses his second best carrier. Thus he cannot gain by changing the carrier from $\widehat{k}$. Similar arguments imply that a small cell $f$ whose best carrier is $\widehat{k}$ but who uses carrier $S_f$ instead of it, will not be interested in changing it to $\widehat{k}$.
\end{IEEEproof}

\end{document}